\documentclass[preprintnumbers]{revtex4}

\usepackage[dvips]{graphicx}
\newcommand{\ie}{{\it i.e.}}
\newcommand{\eg}{{\it e.g.}}
\newcommand{\half}  {\frac{1}{2}}
\newcommand{\qu}{{\rm q}}
\newcommand{\qb}{${\rm\bar q}$}
\newcommand{\pvec}{\vec p}
\newcommand{\kvec}{\vec k}
\newcommand{\rvec}{\vec r}
\newcommand{\Rvec}{\vec R}
\newcommand{\ieps}{i\varepsilon}
\renewcommand{\pl}{{||}}
\newcommand{\order}[1]{${ O}\left(#1 \right)$}
\newcommand{\eq}[1]{(\ref{#1})}
\newcommand{\beq}{\begin{equation}}
\newcommand{\eeq}{\end{equation}}
\newcommand{\ket}[1]{\vert{#1}\rangle}
\renewcommand{\bar}[1]{\overline{#1}}
\newcommand{\M}{{\cal M}}
\newcommand{\VEV}[1]{\left\langle{#1}\right\rangle}
\newcommand{\etal}{{\em et al.}}

\begin{document}
\bibliographystyle{revtex}

\preprint{SLAC--PUB--9022}
\preprint{October 2001}

\title{
Perspectives and Challenges for QCD Phenomenology}\thanks{Work
supported by Department of Energy contract  DE--AC03--76SF00515.}

\author{Stanley J. Brodsky}
\email{sjbth@slac.stanford.edu} \affiliation{Stanford Linear
Accelerator Center, 2575 Sand Hill Road, Menlo Park, CA 94025}

\date{\today}

\begin{abstract}

A fundamental understanding of quantum chromodynamics,
particularly at the amplitude level, is essential for progress in
high energy physics.  For example, the measurement and
interpretation of the basic parameters of the electroweak theory
and $CP$ violation depends on an understanding of the dynamics and
phase structure of exclusive $B$-meson decay amplitudes.
In this review, I discuss a number of ways in which the
required hadron wavefunctions can be measured (such as two-photon
reactions and diffractive dissociation) or calculated from first
principles.  An important tool for describing relativistic
composite systems in quantum field theory is the light-front Fock
expansion, which encodes the properties of a hadrons in terms of a
set of frame-independent $n-$particle wavefunctions.  Light-front
quantization in the doubly-transverse light-cone gauge has a
number of remarkable advantages, including explicit unitarity, the
absence of ghost degrees of freedom, and the decoupling properties
needed to prove factorization theorems in high momentum transfer
inclusive and exclusive reactions.  Evolution in light-cone time
allows the construction of an ``event amplitude generator" in
which only non-ghost physical degrees of freedom and integration
over physical phase appear.  The diffractive dissociation of a
hadron at high energies, by either Coulomb or Pomeron exchange,
has a natural description in QCD as the materialization of the
projectile's light-cone wavefunctions; in particular, the
diffractive dissociation of a meson, baryon, or photon into high
transverse momentum jets measures the shape and other features of
the projectile's distribution amplitude.  Diffractive dissociation
can thus test fundamental properties of QCD, including color
transparency and intrinsic charm.  I also review recent work which
shows that the structure functions measured in deep inelastic
lepton scattering are affected by final-state rescattering, thus
modifying their connection with the light-cone probability
distributions.  In particular, the shadowing of nuclear structure
functions is due to destructive interference effects from
leading-twist diffraction of the virtual photon, physics not
included in the nuclear light-cone wavefunctions.

\end{abstract}

\maketitle

\section{Introduction}

Quantum chromodynamics is the bedrock of the Standard Model,
providing a fundamental description of hadron physics in terms of
quark and gluon degrees of freedom.  The theory has been tested
extensively, particularly in inclusive and exclusive processes
involving collisions at large momentum transfer where
factorization theorems and the smallness of the QCD effective
coupling allow perturbative predictions.  QCD is an extraordinarily
complex and rich theory, leading to many remarkable and novel
physical phenomena.  However, continued testing and development of
QCD, particularly at the amplitude level, is crucial for progress
in high energy physics.  For example, the measurement and
interpretation of the basic parameters of electroweak theory and
$CP$ violation depends on an understanding of the dynamics and
phase structure of exclusive $B$-meson decays amplitudes and the
contributing hadronic wavefunctions.

Despite its empirical successes, many fundamental questions about
QCD have not been resolved.  These include a fundamental
understanding of hadronization and color confinement, the behavior
of the QCD coupling at small momenta, the problem of asymptotic
$n!$ growth of the perturbation theory (renormalon phenomena), the
nature of diffractive phenomena, a fundamental theory of the soft
and hard aspects of the pomeron in high energy reactions, the
origin of shadowing and anti-shadowing in nuclear collisions, the
apparent conflict between QCD vacuum structure and the small size
of the cosmological constant.  There are also a number of
empirical puzzles, such as the anomalous size of the $b \bar b$
production cross section at hadron colliders, the $J/\psi \to \rho
\pi$ puzzle, the apparent small size of spin in the proton, the
strong spin correlations in large angle proton-proton elastic
scattering, the momentum spectrum of $J/\psi$ in $B$ decays, the
unusual pattern of color transparency effects in quasi-elastic
reactions, and the anomalous nuclear dependence of nuclear
structure functions at small momentum transfer.

\section{Light-Front Wavefunctions}

One of the important theoretical goals in QCD is a
frame-independent, quantum-mechanical representation of hadrons at
the amplitude level capable of encoding multi-quark, hidden-color
and gluon momentum, helicity, and flavor correlations in the form
of universal process-independent hadron wavefunctions.  Light-front
quantization allows a unified relativistic wavefunction
representation of non-perturbative hadron dynamics in QCD.
Furthermore, it is possible to measure the wavefunctions of a
relativistic hadron by diffractively dissociating it into jets
whose momentum distribution is correlated with the valence quarks'
momenta
\cite{Ashery:1999nq,Bertsch:1981py,Frankfurt:1993it,Frankfurt:2000tq}.
It is particularly important to understand the shape of the gauge-
and process-independent meson and baryon valence-quark
distribution amplitudes \cite{Lepage:1980fj} $\phi_M(x,Q)$, and
$\phi_B(x_i,Q)$.  These quantities specify how a hadron shares its
longitudinal momentum among its valence quarks; they control
virtually all exclusive processes involving a hard scale $Q$,
including form factors, Compton scattering and photoproduction at
large momentum transfer, as well as the decay of a heavy hadron
into specific final states \cite{Beneke:1999br,Keum:2000ph}.

In light-front quantization, one takes the light-cone time
variable $ t + z/c$ as the evolution parameter instead of ordinary
time $t$.  (The $\hat z$ direction is an arbitrary reference
direction.) The method is often called ``light-front" quantization
rather than ``light-cone" quantization since the equation $x^+ =
\tau = 0$ defines a hyperplane corresponding to a light-front.  The
light-front fixes the initial boundary conditions of a composite
system as its constituents are intercepted by a light-wave
evaluated at a specific value of $x^+ = t + z/c$.  In contrast,
determining an atomic wavefunction at a given instant $t = t_0$
requires measuring the simultaneous scattering of $Z$ photons on
the $Z$ electrons.  An extensive review and guide to the
light-front quantization literature can be found in Ref.
\cite{Brodsky:1998de}.  I will use here the notation $A^\mu =
(A^+, A-, A_\perp),$ where $A^\pm = A^0 \pm A^z$ and the metric $A
\cdot B = {1\over 2 }( A^+ B^- + A^- B^+ ) - A_\perp \cdot
B_\perp.$

It is convenient to define the invariant light-front
Hamiltonian: $ H^{QCD}_{LC} = P^+ P^- - {\vec P_\perp}^2$ where
$P^\pm = P^0 \pm P^z$.  The operator
$P^- = i {d\over d\tau}$ generates light-cone time translations.
The
$P^+$ and
$\vec P_\perp$ momentum operators are independent of
the interactions, and thus are conserved at all orders.
The eigen-spectrum of $ H^{QCD}_{LC}$ in principle gives the entire mass
squared spectrum of color-singlet hadron states in QCD, together with
their respective light-front wavefunctions.  For example, the
proton state satisfies:
$ H^{QCD}_{LC} \ket{\Psi_p} = M^2_p \ket{\Psi_p}$.
The projection of
the proton's eigensolution $\ket{\Psi_p}$ on the color-singlet
$B = 1$, $Q = 1$ eigenstates $\{\ket{n} \}$
of the free Hamiltonian $ H^{QCD}_{LC}(g = 0)$ gives the
light-front Fock expansion: \cite{Brodsky:1989pv}
\begin{eqnarray}
\left\vert \Psi_p; P^+, {\vec P_\perp}, \lambda \right> &=&
\sum_{n \ge 3,\lambda_i}  \int \Pi^{n}_{i=1} {d^2k_{\perp i} dx_i
\over \sqrt{x_i} 16 \pi^3}  16 \pi^3 \delta\left(1- \sum^n_j
x_j\right) \delta^{(2)} \left(\sum^n_\ell \vec k_{\perp
\ell}\right) \\ &&\times \left\vert n; x_i P^+, x_i {\vec P_\perp}
+ {\vec k_{\perp i}}, \lambda_i\right > \psi_{n/p}(x_i,{\vec
k_{\perp i}},\lambda_i) \ .\nonumber
\end{eqnarray}

The light-front Fock wavefunctions $\psi_{n/H}(x_i,{\vec k_{\perp
i}},\lambda_i)$ interpolate between the hadron $H$ and its quark
and gluon degrees of freedom.  The light-cone momentum fractions of
the constituents, $x_i = k^+_i/P^+$ with $\sum^n_{i=1} x_i = 1,$
and the transverse momenta ${\vec k_{\perp i}}$ with $\sum^n_{i=1}
{\vec k_{\perp i}} = {\vec 0_\perp}$ appear as the momentum
coordinates of the light-front Fock wavefunctions.  A crucial
feature is the frame-independence of the light-front
wavefunctions.  The $x_i$ and $\vec k_{\perp i}$ are relative
coordinates independent of the hadron's momentum $P^\mu$.  The
actual physical transverse momenta are ${\vec p_{\perp i}} = x_i
{\vec P_\perp} + {\vec k_{\perp i}}.$ The $\lambda_i$ label the
light-front spin $S^z$ projections of the quarks and gluons along
the $z$ direction.  The physical gluon polarization vectors
$\epsilon^\mu(k,\ \lambda = \pm 1)$ are specified in light-cone
gauge by the conditions $k \cdot \epsilon = 0,\ \eta \cdot
\epsilon = \epsilon^+ = 0.$ Each light-front Fock wavefunction
satisfies conservation of the $z$ projection of angular momentum:
$ J^z = \sum^n_{i=1} S^z_i + \sum^{n-1}_{j=1} l^z_j \ . $ The sum
over $S^z_i$ represents the contribution of the intrinsic spins of
the $n$ Fock state constituents.  The sum over orbital angular
momenta $l^z_j = -{\mathrm i} (k^1_j\frac{\partial}{\partial
k^2_j} -k^2_j\frac{\partial}{\partial k^1_j})$ derives from the
$n-1$ relative momenta.  This excludes the contribution to the
orbital angular momentum due to the motion of the center of mass,
which is not an intrinsic property of the hadron
\cite{Brodsky:2001ii}.

Light-cone wavefunctions represent the ensemble of states
possible when the hadron is intercepted by a light-front at fixed
$\tau = t+z/c.$ The light-cone representation thus provide a
frame-independent, quantum-mechanical representation of the
incoming hadron at the amplitude level, capable of encoding its
multi-quark, hidden-color and gluon momentum, helicity, and flavor
correlations in the form of universal process-independent hadron
wavefunctions.

It is especially convenient to develop the light-front formalism in the
light-cone gauge $A^+ = A^0 + A^z = 0$.  In this gauge the $A^-$ field
becomes a dependent degree of freedom, and it can be eliminated from the
gauge theory Hamiltonian, with the addition of a set of
specific instantaneous light-cone time
interactions.  In fact in $QCD(1+1)$ theory, this instantaneous
interaction provides the confining linear $x^-$ interaction between
quarks.  In $3+1$ dimensions, the transverse field $A^\perp$ propagates
massless spin-one gluon quanta with two polarization vectors
\cite{Lepage:1980fj} which satisfy both the gauge condition
$\epsilon^+_\lambda = 0$ and the Lorentz condition $k\cdot \epsilon= 0$.
Thus no extra condition on the Hilbert space is required
\cite{Srivastava:2000cf}.

There are a number of other simplifications of the light-front
formalism:

\begin{enumerate}
\item
The light-front wavefunctions describe quanta which have positive
energy, positive norm, and physical polarization.  The formalism
is thus physical, and unitary.  No ghosts fields appear
explicitly, even in non-Abelian theory.  The wavefunctions are
only functions of three rather than four physical momentum
variables: the light-front momentum fractions $x_i$ and transverse
momenta $k_\perp$.  The quarks and gluons each have two physical
polarization states.  The
Ward identities for vertex and wavefunction renormalization are
simple for these physical quanta.

\item
The set of light-front wavefunctions provide a frame-independent,
quantum-mechanical description of hadrons at the amplitude level
capable of encoding multi-quark and gluon momentum, helicity, and
flavor correlations in the form of universal process-independent
hadron wavefunctions.  Matrix elements of spacelike currents such
as the spacelike electromagnetic form factors have an exact
representation in terms of simple overlaps of the light-front
wavefunctions in momentum space with the same $x_i$ and unchanged
parton number \cite{Drell:1970km,West:1970av,Brodsky:1980zm}.  In
the case of timelike decays, such as those determined by
semileptonic $B$ decay, one needs to include contributions in
which the parton number $\Delta n =2$ \cite{Brodsky:1999hn}.  The
leading-twist off-forward parton distributions measured in deeply
virtual Compton scattering have a similar light-front wavefunction
representation \cite{Brodsky:2000xy,Diehl:2000xz}.

\item
The high $x \to 1$ and high $k_\perp$ limits of the hadron
wavefunctions control processes and reactions in which the hadron
wavefunctions are highly stressed.  Such configurations involve
far-off-shell intermediate states and can be systematically
treated in perturbation theory
\cite{Brodsky:1995kg,Lepage:1980fj}.

\item
The leading-twist structure functions $q_i(x,Q)$ and $g(x,Q)$
measured in deep inelastic scattering can be computed from the
absolute squares of the light-front wavefunctions, integrated over
the transverse momentum up to the resolution scale $Q$.  All
helicity distributions are thus encoded in terms of the
light-front wavefunctions.  The DGLAP evolution of the structure
functions can be derived from the high $k_\perp$ properties of the
light-front wavefunctions.  Thus given the light-front
wavefunctions, one can compute \cite{Lepage:1980fj} all of the
leading twist helicity and transversity distributions measured in
polarized deep inelastic lepton scattering.  For example, the
helicity-specific quark distributions at resolution $\Lambda$
correspond to
\begin{eqnarray}
q_{\lambda_q/\Lambda_p}(x, \Lambda)
&=& \sum_{n,q_a}
\int\prod^n_{j=1} {dx_j d^2 k_{\perp j}\over 16 \pi^3} \sum_{\lambda_i}
\vert \psi^{(\Lambda)}_{n/H}(x_i,\vec k_{\perp i},\lambda_i)\vert^2
\\
&& \times 16 \pi^3 \delta\left(1- \sum^n_i x_i\right) \delta^{(2)}
\left(\sum^n_i \vec k_{\perp i}\right)
\delta(x - x_q) \delta_{\lambda, \lambda_q}
\Theta(\Lambda^2 - {\cal M}^2_n)\ , \nonumber
\end{eqnarray}
where the sum is over all quarks $q_a$ which match the quantum
numbers, light-front momentum fraction $x,$ and helicity of the
struck quark.  Similarly, the transversity distributions and
off-diagonal helicity convolutions are defined as a density matrix
of the light-front wavefunctions.  This defines the LC
factorization scheme \cite{Lepage:1980fj} where the invariant mass
squared ${\cal M}^2_n = \sum_{i = 1}^n {(k_{\perp i}^2 + m_i^2 )/
x_i}$ of the $n$ partons of the light-front wavefunctions is
limited to $ {\cal M}^2_n < \Lambda^2$.

\item
The distribution of spectator particles in the final state in the
proton fragmentation region in deep inelastic scattering at an
electron-proton collider are encoded in the light-front
wavefunctions of the target proton.  Conversely, the light-front
wavefunctions can be used to describe the coalescence of comoving
quarks into final state hadrons.

\item
The light-front wavefunctions also specify the multi-quark and
gluon correlations of the hadron.  Despite the many sources of
power-law corrections to the deep inelastic cross section, certain
types of dynamical contributions will stand out at large $x_{bj}$
since they arise from compact, highly-correlated fluctuations of
the proton wavefunction.  In particular, there are particularly
interesting dynamical ${\cal O}(1/Q^2)$ corrections which are due
to the {\it interference} of quark currents; {\it i.e.},
contributions which involve leptons scattering amplitudes from two
different quarks of the target nucleon \cite{Brodsky:2000zu}.

\item
The higher Fock states of the light hadrons describe the sea quark
structure of the deep inelastic structure functions, including
``intrinsic" strange\-ness and charm fluctuations specific to the
hadron's structure rather than gluon substructure
\cite{Brodsky:1980pb,Harris:1996jx}.  Ladder relations connecting
state of different particle number follow from the QCD equation of
motion and lead to Regge behavior of the quark and gluon
distributions at $x \to 0$ \cite{Antonuccio:1997tw}.

\item
The gauge- and process-independent meson and baryon valence-quark
distribution amplitudes $\phi_M(x,Q)$, and $\phi_B(x_i,Q)$ which
control exclusive processes involving a hard scale $Q$, including
heavy quark decays, are given by the valence light-front Fock
state wavefunctions integrated over the transverse momentum up to
the resolution scale $Q$.  The evolution equations for
distribution amplitudes follow from the perturbative high
transverse momentum behavior of the light-front wavefunctions
\cite{Brodsky:1989pv}.

\item
The line-integrals needed to defining distribution amplitudes and
structure functions as gauge invariant matrix elements of operator
products vanish in light-front gauge.

\item
Proofs of factorization theorems in hard exclusive and inclusive
reactions are greatly simplified since the propagating gluons in
light-cone gauge couple only to transverse currents; collinear
divergences are thus automatically suppressed.

\item
At high energies each light-front Fock state interacts distinctly;
\eg, Fock states with small particle number and small impact
separation have small color dipole moments and can traverse a
nucleus with minimal interactions.  This is the basis for the
predictions for ``color transparency" in hard quasi-exclusive
\cite{Brodsky:1988xz,Frankfurt:1988nt} and diffractive reactions
\cite{Bertsch:1981py,Frankfurt:1993it,Frankfurt:2000tq}.

\item
The Fock state wavefunctions of hadron can be resolved by a high
energy diffractive interaction, producing forward jets with
momenta which follow the light-front momenta of the wavefunction
\cite{Bertsch:1981py,Frankfurt:1993it,Frankfurt:2000tq}.

\item
 The deuteron form factor at high $Q^2$ is sensitive to
wavefunction configurations where all six quarks overlap within an
impact separation $b_{\perp i} < {\cal O} (1/Q).$ The leading
power-law fall off predicted by QCD is $F_d(Q^2) =
f(\alpha_s(Q^2))/(Q^2)^5$, where, asymptotically,
$f(\alpha_s(Q^2)) \propto \alpha_s(Q^2)^{5+2\gamma}$
\cite{Brodsky:1976rz,Brodsky:1983vf}.  In general, the six-quark
wavefunction of a deuteron is a mixture of five different
color-singlet states.  The dominant color configuration at large
distances corresponds to the usual proton-neutron bound state.
However at small impact space separation, all five Fock
color-singlet components eventually evolve to a state with equal
weight, \ie, the deuteron wavefunction evolves to 80\%\ ``hidden
color'' \cite{Brodsky:1983vf}.  The relatively large normalization
of the deuteron form factor observed at large $Q^2$ hints at
sizable hidden-color contributions \cite{Farrar:1991qi}.  Hidden
color components can also play a predominant role in the reaction
$\gamma d \to J/\psi p n$ at threshold if it is dominated by the
multi-fusion process $\gamma g g \to J/\psi$
\cite{Brodsky:2000zc}.  Hard exclusive nuclear processes can also
be analyzed in terms of ``reduced amplitudes" which remove the
effects of nucleon substructure.
\end{enumerate}

\section{Event Amplitude Generator}

The light-cone formalism can provide the foundations for an
``event amplitude generator" where each quark and gluon final
state is completely labelled in momenta, helicity, and phase.  The
basic idea is to use the light-cone Hamiltonian $P^-$ to generate
the $T$-matrix in light-cone time-ordered perturbation theory in
light-cone gauge.  Loop integrals are integrations over the momenta
of physical quanta and physical phase space $\prod d^2k_{\perp i}
d k^+_i$.  The renormalized amplitudes can be explicitly
constructed by subtracting from the divergent loops amplitudes
with nearly identical integrands corresponding to the contribution
of the relevant mass and coupling counter terms (the ``alternating
denominator method") \cite{Brodsky:1973kb}.  The natural
renormalization scheme to use for defining the coupling in the
event amplitude generator is a physical effective charge such as
the pinch scheme \cite{Cornwall:1989gv}.  The argument of the
coupling is unambiguous.  The DLCQ boundary conditions can be used
to discretized phase space and limit the number of contributing
intermediate states without violating Lorentz invariance.
Hadronization processes can be conceivably incorporated by
convolution with light-cone wavefunctions.  Since one avoids
dimensional regularization and nonphysical ghost degrees of
freedom, this method of generating events at the amplitude level
could be a very simple but powerful tool for simulating events both
in QCD and the Standard Model.

\section{Other Theoretical Tools}

In addition to the light-front Fock expansion, a number of other
useful theoretical tools are available to eliminate theoretical
ambiguities in QCD predictions:

\begin{enumerate}
\item
Conformal symmetry provides a template for QCD predictions
\cite{Brodsky:1999gm}, leading to relations between observables
which are present even in a theory which is not scale invariant.
For example, the natural representation of distribution amplitudes
is in terms of an expansion of orthogonal conformal functions
multiplied by anomalous dimensions determined by QCD evolution
equations \cite{Brodsky:1980ny,Muller:1994hg,Braun:1999te}.  Thus
an important guide in QCD analyses is to identify the underlying
conformal relations of QCD which are manifest if we drop quark
masses and effects due to the running of the QCD couplings.  In
fact, if QCD has an infrared fixed point (vanishing of the
Gell-Mann-Low function at low momenta), the theory will closely
resemble a scale-free conformally symmetric theory in many
applications.

\item
Commensurate scale relations \cite{Brodsky:1995eh,Brodsky:1998ua}
are perturbative QCD predictions which relate observable to
observable at fixed relative scale, such as the ``generalized
Crewther relation" \cite{Brodsky:1996tb}, which connects the
Bjorken and Gross-Llewellyn Smith deep inelastic scattering sum
rules to measurements of the $e^+ e^-$ annihilation cross section.
Such relations have no renormalization scale or scheme ambiguity.
The coefficients in the perturbative series for commensurate scale
relations are identical to those of conformal QCD; thus no
infrared renormalons are present \cite{Brodsky:1999gm}.  One can
identify the required conformal coefficients at any finite order
by expanding the coefficients of the usual PQCD expansion around a
formal infrared fixed point, as in the Banks-Zak method
\cite{Brodsky:2000cr}.  All non-conformal effects are absorbed by
fixing the ratio of the respective momentum transfer and energy
scales.  In the case of fixed-point theories, commensurate scale
relations relate both the ratio of couplings and the ratio of
scales as the fixed point is approached
 \cite{Brodsky:1999gm}.

\item
$\alpha_V$ and Skeleton Schemes.  A physically natural scheme for
defining the QCD coupling in exclusive and other processes is the
$\alpha_V(Q^2)$ scheme defined from the potential of static heavy
quarks.  Heavy-quark lattice gauge theory can provide highly
precise values for the coupling.  All vacuum polarization
corrections due to fermion pairs are then automatically and
analytically incorporated into the Gell-Mann-Low function, thus
avoiding the problem of explicitly computing and resumming quark
mass corrections related to the running of the coupling
\cite{Brodsky:1998mf}.  The use of a finite effective charge such
as $\alpha_V$ as the expansion parameter also provides a basis for
regulating the infrared nonperturbative domain of the QCD
coupling.  A similar coupling and scheme can be based on an
assumed skeleton expansion of the theory
\cite{Cornwall:1989gv,Brodsky:2000cr}.

\item
The Abelian Correspondence Principle.  One can consider QCD
predictions as analytic functions of the number of colors $N_C$
and flavors $N_F$.  In particular, one can show at all orders of
perturbation theory that PQCD predictions reduce to those of an
Abelian theory at $N_C \to 0$ with ${\widehat \alpha} = C_F
\alpha_s$ and ${\widehat N_F} = 2 N_F/C_F$ held fixed
\cite{Brodsky:1997jk}.  There is thus a deep connection between QCD
processes and their corresponding QED analogs.
\end{enumerate}

\section{Other Applications of Light-Front Wavefunctions}

Exclusive semileptonic $B$-decay amplitudes such as $B\rightarrow
A \ell \bar{\nu}$ can also be evaluated exactly in the light-front
formalism \cite{Brodsky:1999hn}.  The time-like decay matrix
elements require the computation of the diagonal matrix element $n
\rightarrow n$ where parton number is conserved, and the
off-diagonal $n+1\rightarrow n-1$ convolution where the current
operator annihilates a $q{\bar{q'}}$ pair in the initial $B$
wavefunction.  This term is a consequence of the fact that the
time-like decay $q^2 = (p_\ell + p_{\bar{\nu}} )^2 > 0$ requires a
positive light-front momentum fraction $q^+ > 0$.  Conversely for
space-like currents, one can choose $q^+=0$, as in the
Drell-Yan-West representation of the space-like electromagnetic
form factors.  However, as can be seen from the explicit analysis
of the form factor in a perturbative model, the off-diagonal
convolution can yield a nonzero $q^+/q^+$ limiting form as $q^+
\rightarrow 0$.  This extra term appears specifically in the case
of ``bad" currents such as $J^-$ in which the coupling to $q\bar
q$ fluctuations in the light-front wavefunctions are favored.  In
effect, the $q^+ \rightarrow 0$ limit generates $\delta(x)$
contributions as residues of the $n+1\rightarrow n-1$
contributions.  The necessity for such ``zero mode" $\delta(x)$
terms was first noted by Chang, Root and Yan \cite{Chang:1973xt},
Burkardt \cite{Burkardt:1989wy}, and Ji and Choi
\cite{Choi:1998nf}.

The off-diagonal $n+1 \rightarrow n-1$ contributions give a new
perspective for the physics of $B$-decays.  A semileptonic decay
involves not only matrix elements where a quark changes flavor, but also
a contribution where the leptonic pair is created from the annihilation
of a $q {\bar{q'}}$ pair within the Fock states of the initial $B$
wavefunction.  The semileptonic decay thus can occur from the
annihilation of a nonvalence quark-antiquark pair in the initial hadron.
This feature will carry over to exclusive hadronic $B$-decays, such as
$B^0 \rightarrow \pi^-D^+$.  In this case the pion can be produced from
the coalescence of a $d\bar u$ pair emerging from the initial higher
particle number Fock wavefunction of the $B$.  The $D$ meson is then
formed from the remaining quarks after the internal exchange of a $W$
boson.

In principle, a precise evaluation of the hadronic matrix elements
needed for $B$-decays and other exclusive electroweak decay amplitudes
requires knowledge of all of the light-front Fock wavefunctions of the
initial and final state hadrons.  In the case of model gauge theories
such as QCD(1+1) \cite{Hornbostel:1990fb} or
collinear QCD \cite{Antonuccio:1995fs} in one-space and one-time dimensions,
the complete
evaluation of the light-front wavefunction is possible for each baryon or
meson bound-state using the DLCQ method.

The virtual Compton scattering process ${d\sigma\over dt}(\gamma^*
p \to \gamma p)$ for large initial photon virtuality
$q^2=-Q^2$ has extraordinary sensitivity to
fundamental features of the proton's structure.  Even though the final
state photon is on-shell, the deeply virtual process probes the
elementary quark structure of the proton near the light cone as an
effective local current.  In contrast to deep inelastic scattering, which
measures only the absorptive part of the forward virtual Compton
amplitude $Im {\cal T}_{\gamma^* p \to
\gamma^* p}$, deeply virtual Compton scattering allows the measurement of
the phase and spin structure of proton matrix elements for general
momentum transfer squared $t$.  In addition, the interference of the
virtual Compton amplitude and Bethe-Heitler
wide angle scattering Bremsstrahlung amplitude where the photon is
emitted from the lepton line leads to an electron-positron asymmetry in
the ${e^\pm  p \to e^\pm \gamma p}$ cross section which is proportional
to the real part of the Compton
amplitude \cite{Brodsky:1972zh,Brodsky:1972vv,Brodsky:1973hm}.  The deeply
virtual Compton amplitude
$\gamma^* p \to \gamma p$ is related by crossing to another important
process
$\gamma^* \gamma
\to $ hadron pairs at fixed invariant mass which can be measured
in electron-photon collisions \cite{Diehl:2000uv}.

In the handbag approximation, the deeply virtual Compton
scattering amplitude $\gamma^*(q) p(P) \to \gamma(q') p(P')$
factorizes as the convolution in $x$ of the amplitude $t^{\mu
\nu}$ for hard Compton scattering on a quark line with the
generalized Compton form factors $H(x,t,\zeta),$ $ E(x,t,\zeta)$,
$\widetilde H(x,t,\zeta),$ and $\widetilde E(x,t,\zeta)$ of the
target
proton \cite{Ji:1996kb,Ji:1997ek,Radyushkin:1996nd,Ji:1998xh,%
Guichon:1998xv,Vanderhaeghen:1998uc,Radyushkin:1999es,%
Collins:1999be,Diehl:1999tr,Diehl:1999kh,Blumlein:2000cx,Penttinen:2000dg}.
Here
$x$ is the light-front momentum fraction of the struck quark, and
$\zeta= Q^2/2 P\cdot q$ plays the role of the Bjorken variable.  The
square of the four-momentum transfer from the proton is given by
$t=\Delta^2\ = \ 2P\cdot \Delta\  =\
-{(\zeta^2M^2+{\vec \Delta_\perp}^2)\over (1-\zeta)}\ $ ,
where $\Delta $ is the difference of initial and final momenta of the
proton ($P=P'+\Delta$).
We will be interested in deeply virtual Compton scattering where $q^2$ is
large compared to the masses and $t$.  Then, to leading order in $1/Q^2$,
${-q^2\over 2P_I\cdot q}=\zeta\ .$
Thus $\zeta$ plays the role of the Bjorken variable in deeply virtual
Compton scattering.
For a fixed value of $-t$, the allowed range of $\zeta$ is given by
\begin{equation}
0\ \le\ \zeta\ \le\
{(-t)\over 2M^2}\ \ \left( {\sqrt{1+{4M^2\over (-t)}}}\ -\ 1\ \right)\ .
\label{nn4}
\end{equation}
The form factor $H(x,t,\zeta)$ describes the proton response when
the helicity of the proton is unchanged, and $E(x,t,\zeta)$ is for
the case when the proton helicity is flipped.  Two additional
functions $\widetilde H(x,t,\zeta),$ and $\widetilde E(x,t,\zeta)$
appear, corresponding to the dependence of the Compton amplitude
on quark helicity.

Recently, Markus Diehl, Dae Sung Hwang and I \cite{Brodsky:2000xy} have
shown how the deeply virtual Compton amplitude can be evaluated
explicitly in the Fock state representation using the matrix elements of
the currents and the boost properties of the light-front wavefunctions.
For the $n \to n$ diagonal term ($\Delta n = 0$), the arguments of the
final-state hadron wavefunction are
$x_1-\zeta \over 1-\zeta$,
${\vec{k}}_{\perp 1} - {1-x_1\over 1-\zeta} {\vec{\Delta}}_\perp$ for
the struck quark
and $x_i\over 1-\zeta$,
${\vec{k}}_{\perp i} + {x_i\over 1-\zeta} {\vec{\Delta}}_\perp$
for the $n-1$ spectators.  As in the case of leptonic $B$ decays, one also
the evaluation of an $n+1 \to n-1$ off-diagonal term ($\Delta n = -2$),
where
partons $1$ and
$n+1$ of the initial wavefunction annihilate into the current leaving
$n-1$ spectators.
Then $x_{n+1} = \zeta - x_{1}$,
${\vec{k}}_{\perp n+1} = {\vec{\Delta}}_\perp-{\vec{k}}_{\perp 1}$.
The remaining $n-1$ partons have total momentum
$((1-\zeta)P^+, -{\vec{\Delta}}_{\perp})$.
The final wavefunction then has arguments
$x^\prime_i = {x_i \over 1- \zeta}$ and
${\vec{k}}^\prime_{\perp i}=
{\vec{k}}_{\perp i} + {x_i\over 1-\zeta} {\vec{\Delta}}_\perp$.

\section{Applications of QCD Factorization to Hard QCD
 Processes}

Factorization theorems for hard exclusive, semi-exclusive, and
diffractive processes allow the separation of soft
non-perturbative dynamics of the bound state hadrons from the hard
dynamics of a perturbatively-calculable quark-gluon scattering
amplitude.  The factorization of inclusive reactions is reviewed in ref.
For reviews and bibliography of exclusive process calculations in QCD
(see Ref. \cite{Brodsky:1989pv,Brodsky:2000dr}).

The light-front
formalism provides a physical factorization scheme which conveniently
separates and factorizes soft non-perturbative physics from hard
perturbative dynamics in both exclusive and
inclusive reactions \cite{Lepage:1980fj,Lepage:1979zb}.

In hard inclusive
reactions all intermediate states are divided according to $\M^2_n <
\Lambda^2 $ and
$\M^2_n >
\Lambda^2 $ domains.  The lower mass regime is associated with the quark
and gluon distributions defined from the absolute squares of the LC
wavefunctions in the light cone factorization scheme.  In the high
invariant mass regime, intrinsic transverse momenta can be ignored, so
that the structure of the process at leading power has the form of hard
scattering on collinear quark and gluon constituents, as in the parton
model.  The attachment of gluons from the LC wavefunction to a propagator
in a hard subprocess is power-law suppressed in LC gauge, so that the
minimal quark-gluon particle-number subprocesses dominate.  It is then
straightforward to derive the DGLAP equations from the evolution of the
distributions with $\log \Lambda^2$.
The
anomaly contribution to singlet helicity structure function $g_1(x,Q)$
can be explicitly identified in the LC factorization scheme as due to the
$\gamma^* g \to q
\bar q$ fusion process.  The anomaly contribution would be zero if the
gluon is on shell.  However, if the off-shellness of the state is larger
than the quark pair mass, one obtains the usual anomaly
contribution \cite{Bass:1999rn}.

In exclusive amplitudes, the LC wavefunctions are the interpolating
amplitudes connecting the quark and gluons to the hadronic
states.  In an
exclusive amplitude involving a hard scale $Q^2$ all intermediate states
can be divided according to
$\M^2_n <
\Lambda^2 < Q^2 $ and $\M^2_n < \Lambda^2 $ invariant mass domains.  The
high invariant mass contributions to the amplitude has the structure of a
hard scattering process
$T_H$ in which the hadrons are replaced by their respective (collinear)
quarks and gluons.  In light-cone gauge only the minimal Fock states
contribute to the leading power-law fall-off of the exclusive amplitude.
The wavefunctions in the lower invariant mass domain can be integrated up
to an arbitrary intermediate invariant mass cutoff $\Lambda$.  The
invariant mass domain beyond this cutoff is included in the hard
scattering amplitude
$T_H$.  The $T_H$ satisfy dimensional
counting rules \cite{Brodsky:1975vy}.  Final-state and initial state
corrections from gluon attachments to lines connected to the
color-singlet distribution amplitudes cancel at leading twist.  Explicit
examples of perturbative QCD factorization will be discussed in more
detail in the next section.

The key non-perturbative input for exclusive
processes is thus the gauge and frame independent hadron distribution
amplitude \cite{Lepage:1979zb,Lepage:1980fj} defined as the integral of
the valence (lowest particle number) Fock wavefunction;
\eg\ for the pion
\begin{equation}
\phi_\pi (x_i,\Lambda) \equiv \int d^2k_\perp\, \psi^{(\Lambda)}_{q\bar
q/\pi} (x_i, \vec k_{\perp i},\lambda)
\label{eq:f1a}
\end{equation}
where the global cutoff $\Lambda$ is identified with the
resolution $Q$.  The distribution amplitude controls leading-twist
exclusive amplitudes at high momentum transfer, and it can be
related to the gauge-invariant Bethe-Salpeter wavefunction at
equal light-cone time.  The logarithmic evolution of hadron
distribution amplitudes $\phi_H (x_i,Q)$ can be derived from the
perturbatively-computable tail of the valence light-front
wavefunction in the high transverse momentum regime
\cite{Lepage:1979zb,Lepage:1980fj}.  The conformal basis for the
evolution of the three-quark distribution amplitudes for the
baryons~ \cite{Lepage:1979za} has recently been obtained by V.
Braun \etal \ \cite{Braun:1999te}

The existence of an exact formalism
provides a basis for systematic approximations and a control over neglected
terms.  For example, one can analyze exclusive semi-leptonic
$B$-decays which involve hard internal momentum transfer using a
perturbative QCD
formalism \cite
{Szczepaniak:1990dt,Szczepaniak:1996xg,%
Beneke:1999br,Keum:2000ph,Keum:2000wi,Li:2000hh}
patterned after the perturbative analysis of form
factors at large momentum transfer.  The hard-scattering
analysis again proceeds
by writing each hadronic wavefunction
as a sum of soft and hard contributions
\begin{equation}
\psi_n = \psi^{{\rm soft}}_n (\M^2_n < \Lambda^2) + \psi^{{\rm hard}}_n
(\M^2_n >\Lambda^2) ,
\end{equation}
where $\M^2_n $ is the invariant mass of the partons in the $n$-particle
Fock state and
$\Lambda$ is the separation scale.
The high internal momentum contributions to the wavefunction $\psi^{{\rm
hard}}_n $ can be calculated systematically from QCD perturbation theory
by iterating the gluon exchange kernel.  The contributions from
high momentum transfer exchange to the
$B$-decay amplitude can then be written as a
convolution of a hard-scattering
quark-gluon scattering amplitude $T_H$ with the distribution
amplitudes $\phi(x_i,\Lambda)$, the valence wavefunctions obtained by
integrating the
constituent momenta up to the separation scale
${\cal M}_n < \Lambda < Q$.  Furthermore in processes such as $B \to \pi
D$ where the pion is effectively produced as a rapidly-moving small Fock
state with a small color-dipole interactions,  final state interactions
are suppressed by color transparency.  This is the basis for the
perturbative hard-scattering
analyses \cite{Szczepaniak:1990dt,Beneke:1999br,Keum:2000ph,%
Keum:2000wi,Li:2000hh}.
In a systematic
analysis, one can identify the hard PQCD contribution as well as the soft
contribution from the convolution of the light-front wavefunctions.
Furthermore, the hard-scattering contribution can be systematically improved.

Given the solution
for the hadronic wavefunctions $\psi^{(\Lambda)}_n$ with $\M^2_n <
\Lambda^2$, one can construct the wavefunction in the hard regime with
$\M^2_n > \Lambda^2$ using projection operator techniques.  The
construction can be done perturbatively in QCD since only high invariant mass,
far off-shell matrix elements are involved.  One can use this method to
derive the physical properties of the LC wavefunctions and their matrix elements
at high invariant mass.  Since $\M^2_n = \sum^n_{i=1}
\left(\frac{k^2_\perp+m^2}{x}\right)_i $, this method also allows the derivation
of the asymptotic behavior of light-front wavefunctions at large $k_\perp$,
which
in turn leads to predictions for the fall-off of form factors and other
exclusive
matrix elements at large momentum transfer, such as the quark counting rules
for predicting the nominal power-law fall-off of two-body scattering amplitudes
at fixed
$\theta_{cm}$ \cite{Brodsky:1975vy} and
helicity selection rules \cite{Brodsky:1981kj}.  The phenomenological
successes of these rules
can be understood within QCD if the coupling
$\alpha_V(Q)$ freezes in a range of relatively
small momentum transfer \cite{Brodsky:1998dh}.

\section{Two-Photon Processes}

The simplest and perhaps the most elegant illustration of an
exclusive reaction in QCD is the evaluation of the photon-to-pion
transition form factor $F_{\gamma \to \pi}(Q^2)$
\cite{Lepage:1980fj,Brodsky:1981rp} which is measurable in
single-tagged two-photon $ee \to ee \pi^0$ reactions.  The form
factor is defined via the invariant amplitude $ \Gamma^\mu = -ie^2
F_{\pi \gamma}(Q^2) \epsilon^{\mu \nu \rho \sigma} p^\pi_\nu
\epsilon_\rho q_\sigma \ .$ As in inclusive reactions, one must
specify a factorization scheme which divides the integration
regions of the loop integrals into hard and soft momenta, compared
to the resolution scale $\widetilde Q$.  At leading twist, the
transition form factor then factorizes as a convolution of the
$\gamma^* \gamma \to q \bar q$ amplitude (where the quarks are
collinear with the final state pion) with the valence light-front
wavefunction of the pion:
\begin{equation}
F_{\gamma M}(Q^2)= {4 \over \sqrt 3}\int^1_0 dx
\phi_M(x,\widetilde Q) T^H_{\gamma \to M}(x,Q^2) .
\label{transitionformfactor}
\end{equation}
The hard scattering amplitude for $\gamma\gamma^*\to q \bar q$ is
$ T^H_{\gamma M}(x,Q^2) = { [(1-x) Q^2]^{-1}}\times \left(1 +
{\cal O}(\alpha_s)\right). $ The leading QCD corrections have been
computed by Braaten \cite{Braaten:1987yy}.  The evaluation of the
next-to-leading corrections in the physical $\alpha_V$ scheme is
given in Ref. \cite{Brodsky:1998dh}. For the asymptotic
distribution amplitude $\phi^{\rm asympt}_\pi (x) = \sqrt 3 f_\pi
x(1-x)$ one predicts $ Q^2 F_{\gamma \pi}(Q^2)= 2 f_\pi \left(1 -
{5\over3} {\alpha_V(Q^*)\over \pi}\right)$ where $Q^*= e^{-3/2} Q$
is the BLM scale for the pion form factor.  The PQCD predictions
have been tested in measurements of $e \gamma \to e \pi^0$ by the
CLEO collaboration \cite{Gronberg:1998fj}.  The observed flat
scaling of the $Q^2 F_{\gamma \pi}(Q^2)$ data from $Q^2 = 2$ to
$Q^2 = 8$ GeV$^2$ provides an important confirmation of the
applicability of leading twist QCD to this process.  The magnitude
of $Q^2 F_{\gamma \pi}(Q^2)$ is remarkably consistent with the
predicted form, assuming the asymptotic distribution amplitude and
including the LO QCD radiative correction with $\alpha_V(e^{-3/2}
Q)/\pi \simeq 0.12$.  One could allow for some broadening of the
distribution amplitude with a corresponding increase in the value
of $\alpha_V$ at small scales.  Radyushkin
\cite{Radyushkin:1995pj}, Ong \cite{Ong:1995gs}, and Kroll
\cite{Kroll:1996jx} have also noted that the scaling and
normalization of the photon-to-pion transition form factor tends
to favor the asymptotic form for the pion distribution amplitude
and rules out broader distributions such as the two-humped form
suggested by QCD sum rules \cite{Chernyak:1984ej}.

The two-photon annihilation process $\gamma^* \gamma \to $
hadrons, which is measurable in single-tagged $e^+ e^- \to e^+ e^- {\rm
hadrons}$ events, provides a semi-local probe of
$C=+$ hadron systems $\pi^0, \eta^0, \eta^\prime, \eta_c, \pi^+ \pi^-$,
etc.  The $\gamma^* \gamma
\to \pi^+
\pi^-$ hadron pair process is related to virtual Compton
scattering on a pion target by crossing.  The leading twist amplitude is
sensitive to the
$1/x - 1/(1-x)$ moment of the two-pion distribution amplitude coupled
to two valence quarks \cite{Muller:1994fv,Diehl:2000uv}.

\begin{figure}[htb]
 \includegraphics{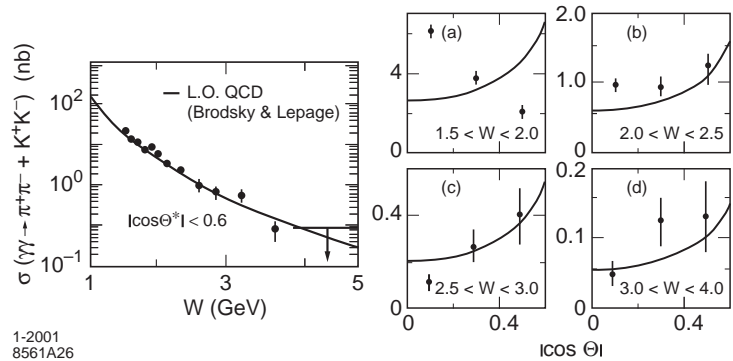}%
 \caption{ \label{Fig:CLEO}
 Comparison of the sum of $\gamma \gamma \rightarrow \pi^+
\pi^-$ and $\gamma \gamma \rightarrow K^+ K^-$ meson pair
production cross sections with the scaling and angular
distribution of the perturbative QCD prediction
\cite{Brodsky:1981rp}.  The data are from the CLEO collaboration
\cite{Paar}.}
 \end{figure}

Two-photon reactions, $\gamma \gamma \to H \bar H$ at large s =
$(k_1 + k_2)^2$ and fixed $\theta_{\rm cm}$, provide a
particularly important laboratory for testing QCD since these
cross-channel ``Compton" processes are the simplest calculable
large-angle exclusive hadronic scattering reactions.  The helicity
structure, and often even the absolute normalization can be
rigorously computed for each two-photon channel
\cite{Brodsky:1981rp}.  In the case of meson pairs, dimensional
counting predicts that for large $s$, $s^4 d\sigma/dt(\gamma
\gamma \to M \bar M$ scales at fixed $t/s$ or $\theta_{\rm c.m.}$
up to factors of $\ln s/\Lambda^2$.  The angular dependence of the
$\gamma \gamma \to H \bar H$ amplitudes can be used to determine
the shape of the process-independent distribution amplitudes,
$\phi_H(x,Q)$.  An important feature of the $\gamma \gamma \to M
\bar M$ amplitude for meson pairs is that the contributions of
Landshoff pitch singularities are power-law suppressed at the Born
level -- even before taking into account Sudakov form factor
suppression.  There are also no anomalous contributions from the
$x \to 1$ endpoint integration region.  Thus, as in the calculation
of the meson form factors, each fixed-angle helicity amplitude can
be written to leading order in $1/Q$ in the factorized form $[Q^2
= p_T^2 = tu/s; \widetilde Q_x = \min(xQ,(l-x)Q)]$:
\begin{equation}{\cal M}_{\gamma \gamma\to M \bar M}
= \int^1_0 dx \int^1_0 dy \phi_{\bar M}(y,\widetilde Q_y)
T_H(x,y,s,\theta_{\rm c.m.} \phi_{M}(x,\widetilde Q_x) ,
\end{equation} where $T_H$ is the hard-scattering amplitude
$\gamma \gamma \to (q \bar q) (q \bar q)$ for the production of
the valence quarks collinear with each meson, and
$\phi_M(x,\widetilde Q)$ is the amplitude for finding the valence
$q$ and $\bar q$ with light-front fractions of the meson's
momentum, integrated over transverse momenta $k_\perp < \widetilde
Q.$ The contribution of non-valence Fock states are power-law
suppressed.  Furthermore, the helicity-selection rules
\cite{Brodsky:1981kj} of perturbative QCD predict that vector
mesons are produced with opposite helicities to leading order in
$1/Q$ and all orders in $\alpha_s$.  The dependence in $x$ and $y$
of several terms in $T_{\lambda, \lambda'}$ is quite similar to
that appearing in the meson's electromagnetic form factor.  Thus
much of the dependence on $\phi_M(x,Q)$ can be eliminated by
expressing it in terms of the meson form factor.  In fact, the
ratio of the $\gamma \gamma \to \pi^+ \pi^-$ and $e^+ e^- \to
\mu^+ \mu^-$ amplitudes at large $s$ and fixed $\theta_{CM}$ is
nearly insensitive to the running coupling and the shape of the
pion distribution amplitude:
\begin{equation}{{d\sigma \over dt }(\gamma \gamma \to \pi^+ \pi^-)
\over {d\sigma \over dt }(\gamma \gamma \to \mu^+ \mu^-)} \sim {4
\vert F_\pi(s) \vert^2 \over 1 - \cos^2 \theta_{\rm c.m.} }
.\end{equation} The comparison of the PQCD prediction for the sum
of $\pi^+ \pi^-$ plus $K^+ K^-$ channels with recent CLEO data
\cite{Paar} is shown in Fig. \ref{Fig:CLEO}.  The CLEO data for
charged pion and kaon pairs show a clear transition to the scaling
and angular distribution predicted by PQCD \cite{Brodsky:1981rp}
for $W = \sqrt(s_{\gamma \gamma} > 2$ GeV.  It is clearly
important to measure the magnitude and angular dependence of the
two-photon production of neutral pions and $\rho^+ \rho^-$ cross
sections in view of the strong sensitivity of these channels to
the shape of meson distribution amplitudes.  QCD also predicts that
the production cross section for charged $\rho$-pairs (with any
helicity) is much larger that for that of neutral $\rho$ pairs,
particularly at large $\theta_{\rm c.m.}$ angles.  Similar
predictions are possible for other helicity-zero mesons.  The cross
sections for Compton scattering on protons and the crossed
reaction $\gamma \gamma \to p \bar p$ at high momentum transfer
have also been evaluated \cite{Farrar:1990qj,Brooks:2000nb},
providing important tests of the proton distribution amplitude.

It is
particularly compelling to see a transition in angular dependence between
the low energy chiral and PQCD regimes.  The success of leading-twist
perturbative QCD scaling
for exclusive processes at presently experimentally accessible momentum
transfer can be understood if the effective coupling
$\alpha_V(Q^*)$ is approximately constant at the relatively
small scales $Q^*$ relevant to the hard scattering
amplitudes \cite{Brodsky:1998dh}.  The evolution of the quark distribution
amplitudes In the low-$Q^*$ domain at also needs to be minimal.  Sudakov
suppression of the endpoint contributions is also strengthened if the
coupling is frozen because of the exponentiation of a double logarithmic
series.

Clearly much more experimental input on hadron wavefunctions is needed,
particularly from measurements of two-photon exclusive reactions into
meson and baryon pairs at the high luminosity
$B$ factories.  For example, the ratio ${{d\sigma \over dt
}(\gamma
\gamma \to \pi^0
\pi^0)
/ {d\sigma \over dt}(\gamma \gamma \to \pi^+ \pi^-)}$
is particularly sensitive to the shape of pion distribution amplitude.
Baryon pair production in two-photon reactions at threshold may reveal
physics associated with the soliton structure of baryons in
QCD \cite{Sommermann:1992yh,Marek:2001}.  In addition, fixed target
experiments can provide much more information on fundamental QCD
processes such as deeply virtual Compton scattering and large angle
Compton scattering.

\section{Diffraction and Light-Cone Wavefunctions}

The diffractive dissociation of a hadron at high energies, by
either Coulomb or Pomeron exchange, can be understood as the
materialization of the projectile's light-cone wavefunctions; in
particular, the diffractive dissociation of a meson, baryon, or
photon into high transverse momentum jets measures the shape and
other features of the projectile's distribution amplitude,
$\phi(x_i,Q),$ the valence wavefunction which controls high
momentum transfer exclusive amplitudes.  Diffractive dissociation
can also test fundamental properties of QCD, including color
transparency and intrinsic charm.

Diffractive dissociation in QCD can be understood as a three-step
process:

\begin{enumerate}
\item
The initial hadron can be decomposed in terms of its quark and
gluon constituents in terms of its light-cone Fock-state
wavefunctions.

\item
In the second step, the incoming hadron is resolved by Pomeron or
Odderon (multi-gluon) exchange with the target or by Coulomb
dissociation.  The exchanged interaction has to supply sufficient
momentum transfer $q^\mu$ to put the diffracted state $X$ on
shell.  Light-cone energy conservation requires $q^- = {( m_X^2-
m_\pi^2)/ P^+_\pi},$ where $m_X$ is the invariant mass of $X$.  In
a heavy target rest system, the longitudinal momentum transfer
for a pion beam is
$q^z = {(m_X^2- m_\pi^2)/ E_{\pi {\rm lab}}}.$ Thus the momentum
transfer $t = q^2$ to the target can be sufficiently small so that
the target remains intact.

In perturbative QCD, the pomeron is generally be represented as
multiple gluon exchange between the target and projectile.
Effectively this interaction occurs over a short light-cone time
interval, and thus like photon exchange, the perturbative QCD
pomeron can be effectively represented as a local operator.  This
description is believed to be applicable when the pomeron has to
resolve compact states and is the basis for the terminology ``hard
pomeron".  The BFKL formalism generalizes the perturbative QCD
treatment by an all-orders perturbative resummation, generating a
pomeron with a fixed Regge intercept $\alpha_P(0)$.  Next to
leading order calculations with BLM scale fixing leads to a
predicted intercept $\alpha_P(0) \simeq
0.4$~\cite{Brodsky:1999kn}.  However, when the exchange
interactions are soft, a multiperipheral description in terms of
meson ladders may dominate the physics.  This is the basis for the
two-component pomeron model of Donnachie and
Landshoff~\cite{Donnachie:2001xx}.

Consider a collinear frame where the incident momentum $P^+_\pi$
is large and $s = (p_\pi + p_{\rm target})^2 \simeq p^+_\pi
p^-_{\rm target}.$ The matrix element of an exchanged gluon with
momentum $q_i$ between the projectile and an intermediate state
$\ket N$ is dominated by the ``plus current": $\VEV{\pi|j^+(0)\exp
(i {\half q^+_i x^-}-i q_{\perp i} \cdot x_\perp|N}$.  Note that
the coherent sum of couplings of an exchanged gluon to the pion
system vanishes when its momentum is small compared to the
characteristic momentum scales in the projectile light-cone
wavefunction: $q^{\perp_i} \Delta x_\perp \ll 1$ and $q^+_i \Delta
x^- \ll 1$.  The destructive interference of the gauge couplings
to the constituents of the projectile follows simply from the fact
that the color charge operator has zero matrix element between
distinct eigenstates of the QCD Hamiltonian: $\VEV{A|Q|B} \equiv
\int d^2x{_\perp} dx^- \VEV{A|j^+(0)|B} = 0$~\cite{BH2}.  At high
energies the change in $k^+_i$ of the constituents can be ignored,
so that Fock states of a hadron with small transverse size
interact weakly even in a nuclear target because of their small
dipole moment.  This is the basis of ``color transparency'' in
perturbative QCD~\cite{Brodsky:1988xz,Bertsch:1981py}.  To a good
approximation the sum of couplings to the constituents of the
projectile can be represented as a derivative with respect to
transverse momentum.  Thus photon exchange measures a weighted sum
of transverse derivatives $\partial_{k_\perp} \psi_n(x_i,
k_{\perp_i},\lambda_i),$ and two-gluon exchange measures the
second transverse partial derivative~\cite{BHDP}.

\item
The final step is the hadronization of the $n$ constituents of the
projectile Fock state into final state hadrons.  Since $q^+_i$ is
small, the number of partons in the initial Fock state and the
final state hadrons are unchanged.  Their coalescence is thus
governed by the convolution of initial and final-state Fock state
wavefunctions.  In the case of states with high $k_\perp$, the
final state will hadronize into jets, each reflecting the
respective $x_i$ of the Fock state constituents.  In the case of
higher Fock states with intrinsic sea quarks such as an extra $c
\bar c$ pair (intrinsic charm), one will observe leading $J/\psi$
or open charm hadrons in the projectile fragmentation region; \ie,
the hadron's fragments will tend to have the same rapidity as that
of the projectile.

\end{enumerate}

For example, diffractive multi-jet production in heavy nuclei
provides a novel way to measure the shape of the LC Fock state
wavefunctions and test color transparency.  Consider the reaction
\cite{Bertsch:1981py,Frankfurt:1993it,Frankfurt:2000jm} $\pi A
\rightarrow {\rm Jet}_1 + {\rm Jet}_2 + A^\prime$ at high energy
where the nucleus $A^\prime$ is left intact in its ground state.
The transverse momenta of the jets balance so that $ \vec k_{\perp
i} + \vec k_{\perp 2} = \vec q_\perp < {R^{-1}}_A \ . $ The
light-front longitudinal momentum fractions also need to add to
$x_1+x_2 \sim 1$ so that $\Delta p_L < R^{-1}_A$.  The process can
then occur coherently in the nucleus.  Because of color
transparency, the valence wavefunction of the pion with small
impact separation, will penetrate the nucleus with minimal
interactions, diffracting into jet pairs \cite{Bertsch:1981py}.
The $x_1=x$, $x_2=1-x$ dependence of the di-jet distributions will
thus reflect the shape of the pion valence light-front
wavefunction in $x$; similarly, the $\vec k_{\perp 1}- \vec
k_{\perp 2}$ relative transverse momenta of the jets gives key
information on the second derivative of the underlying shape of
the valence pion wavefunction
\cite{Frankfurt:1993it,Frankfurt:2000jm,BHDP}.  The diffractive
nuclear amplitude extrapolated to $t = 0$ should be linear in
nuclear number $A$ if color transparency is correct.  The
integrated diffractive rate should then scale as $A^2/R^2_A \sim
A^{4/3}$.

The results of a diffractive dijet dissociation experiment of this
type E791 at Fermilab using 500 GeV incident pions on nuclear
targets \cite{Aitala:2001hc} appear to be consistent with color
transparency.  The measured longitudinal momentum distribution of
the jets \cite{Aitala:2001hb} is consistent with a pion light-cone
wavefunction of the pion with the shape of the asymptotic
distribution amplitude, $\phi^{\rm asympt}_\pi (x) = \sqrt 3 f_\pi
x(1-x)$.  Data from CLEO \cite{Gronberg:1998fj} for the $\gamma
\gamma^* \rightarrow \pi^0$ transition form factor also favor a
form for the pion distribution amplitude close to the asymptotic
solution to the perturbative QCD evolution equation
\cite{Lepage:1980fj}.

The interpretation of the diffractive dijet processes as measures
of the hadron distribution amplitudes has recently been questioned
by Braun {\em et al.} \cite{Braun:2001ih} and by Chernyak
\cite{Chernyak:2001ph} who have calculated the hard scattering
amplitude for such processes at next-to-leading order.  However,
these analyses neglect the integration over the transverse
momentum of the valence quarks and thus miss the logarithmic
ordering which is required for factorization of the distribution
amplitude and color-filtering in nuclear targets.

As noted above, the diffractive dissociation of a hadron or
nucleus can also occur via the Coulomb dissociation of a beam
particle on an electron beam (\eg\ at HERA or eRHIC) or on the
strong Coulomb field of a heavy nucleus (\eg\ at RHIC or nuclear
collisions at the LHC) \cite{BHDP}.  The amplitude for Coulomb
exchange at small momentum transfer is proportional to the first
derivative $\sum_i e_i {\partial \over \vec k_{T i}} \psi$ of the
light-front wavefunction, summed over the charged constituents.
The Coulomb exchange reactions fall off less fast at high
transverse momentum compared to pomeron exchange reactions since
the light-front wavefunction is effective differentiated twice in
two-gluon exchange reactions.

It will also be interesting to study diffractive tri-jet
production using proton beams $ p A \rightarrow {\rm Jet}_1 + {\rm
Jet}_2 + {\rm Jet}_3 + A^\prime $ to determine the fundamental
shape of the 3-quark structure of the valence light-front
wavefunction of the nucleon at small transverse separation
\cite{Frankfurt:1993it}.  For example, consider the Coulomb
dissociation of a high energy proton at HERA.  The proton can
dissociate into three jets corresponding to the three-quark
structure of the valence light-front wavefunction.  We can demand
that the produced hadrons all fall outside an opening angle
$\theta$ in the proton's fragmentation region.  Effectively all of
the light-front momentum $\sum_j x_j \simeq 1$ of the proton's
fragments will thus be produced outside an ``exclusion cone".  This
then limits the invariant mass of the contributing Fock state ${
M}^2_n > \Lambda^2 = P^{+2} \sin^2\theta/4$ from below, so that
perturbative QCD counting rules can predict the fall-off in the
jet system invariant mass $ M$.  The segmentation of the forward
detector in azimuthal angle $\phi$ can be used to identify
structure and correlations associated with the three-quark
light-front wavefunction \cite{BHDP}.  One can use also measure the
dijet structure of real and virtual photons beams $ \gamma^* A
\rightarrow {\rm Jet}_1 + {\rm Jet}_2 + A^\prime $ to measure the
shape of the light-front wavefunction for transversely-polarized
and longitudinally-polarized virtual photons.  Such experiments
will open up a direct window on the amplitude structure of hadrons
at short distances.  The light-front formalism is also applicable
to the description of nuclei in terms of their nucleonic and
mesonic degrees of freedom \cite{Miller:2001mi,Miller:2000ta}.
Self-resolving diffractive jet reactions in high energy
electron-nucleus collisions and hadron-nucleus collisions at
moderate momentum transfers can thus be used to resolve the
light-front wavefunctions of nuclei.

Thus diffractive jet
production can
provide direct empirical information on the light-front wavefunctions of
hadrons.
The E791 experiment at Fermilab has not only determined the main features of
the pion wavefunction, but has also
confirmed color transparency, a fundamental test of the gauge properties of
QCD.
Analogous reaction involving nuclear projectiles can resolve the light-front
wavefunctions of
nuclei in terms of their nucleon and mesonic degrees of freedom.
It is also possible to measure the light-front wavefunctions of atoms through
high energy Coulomb dissociation.

\section{Heavy Quark Fluctuations in Diffractive Dissociation}

Since a hadronic wavefunction describes states off of the
light-cone energy shell, there is a finite probability of the
projectile having fluctuations containing extra quark-antiquark
pairs, such as intrinsic strangeness charm, and bottom.  In
contrast to the quark pairs arising from gluon splitting,
intrinsic quarks are multiply-connected to the valence quarks and
are thus part of the dynamics of the hadron.  Recently Franz,
Polyakov, and Goeke have analyzed the properties of the intrinsic
heavy-quark fluctuations in hadrons using the operator-product
expansion~\cite{Franz:2000ee}.  For example, the light-cone
momentum fraction carried by intrinsic heavy quarks in the proton
$x_{Q \bar Q}$ as measured by the $T^{+ + }$ component of the
energy-momentum tensor is related in the heavy-quark limit to the
forward matrix element $\langle p \vert {\hbox{tr}_c}
{(G^{+\alpha} G^{+ \beta} G_{\alpha \beta})/ m_Q^2 }\vert p
\rangle ,$ where $G^{\mu \nu}$ is the gauge field strength tensor.
Diagrammatically, this can be described as a heavy quark loop in
the proton self-energy with four gluons attached to the light,
valence quarks.  Since the non-Abelian commutator $[A_\alpha,
A_\beta]$ is involved, the heavy quark pairs in the proton
wavefunction are necessarily in a color-octet state.  It follows
from dimensional analysis that the momentum fraction carried by
the $Q\bar Q$ pair scales as $k^2_\perp / m^2_Q$ where $k_\perp$
is the typical momentum in the hadron wave function.  [In
contrast, in the case of Abelian theories, the contribution of an
intrinsic, heavy lepton pair to the bound state's structure first
appears in ${ O}(1/m_L^4)$.  One relevant operator corresponds to
the Born-Infeld $(F_{\mu\nu})^4$ light-by-light scattering
insertion, and the momentum fraction of heavy leptons in an atom
scales as $k^4_\perp / m_L^4$.]

Intrinsic charm can be materialized by diffractive dissociation
into open or hidden charm states such as $p p \to J/\psi X p',
\Lambda_c X p'$.  At HERA one can measure intrinsic charm in the
proton by Coulomb dissociation: $p e \to \Lambda_C X e',$ and
$J/\psi X e'.$ Since the intrinsic heavy quarks tend to have the
same rapidity as that of the projectile, they are produced at
large $x_F$ in the beam fragmentation region.  The charm structure
function measured by the EMC group shows an excess at large
$x_{bj}$, indicating a probability of order $1\%$ for intrinsic
charm in the proton~\cite{Harris:1996jx}.  The presence of
intrinsic charm in light-mesons provides an explanation for the
puzzle of the large $J/\psi \to \rho\pi$ branching ratio and
suppressed $\psi^\prime \to \rho\pi$ decay~\cite{Brodsky:1997fj}.
The presence of intrinsic charm quarks in the $B$ wave function
provides new mechanisms for $B$ decays.  For example, Chang and
Hou have considered the production of final states with three
charmed quarks such as $B \to J/\psi D \pi$ and $B \to J/\psi
D^*$~\cite{Chang:2001iy}; these final states are difficult to
realize in the valence model, yet they occur naturally when the
$b$ quark of the intrinsic charm Fock state $\ket{ b \bar u c \bar
c}$ decays via $b \to c \bar u d$.  In fact, the $J/\psi$ spectrum
for inclusive $B \to J/\psi X$ decays measured by CLEO and Belle
shows a distinct enhancement at the low $J/\psi$ momentum where
such decays would kinematically occur.  Alternatively, this excess
could reflect the opening of baryonic channels such as $B \to
J/\psi \bar p \Lambda$~\cite{Brodsky:1997yr}.  Recently, Susan
Gardner and I have shown that the presence of intrinsic charm in
the hadrons' light-cone wave functions, even at a few percent
level, provides new, competitive decay mechanisms for $B$ decays
which are nominally CKM-suppressed~\cite{Brodsky:2001yt}.  For
example, the weak decays of the $B$-meson to two-body exclusive
states consisting of strange plus light hadrons, such as $B \to
\pi K$, are expected to be dominated by penguin contributions
since the tree-level $b\to s u{\overline u}$ decay is CKM
suppressed.  However, higher Fock states in the $B$ wave function
containing charm quark pairs can mediate the decay via a
CKM-favored $b\to s c{\overline c}$ tree-level transition.  Such
intrinsic charm contributions can be phenomenologically
significant.  Since they mimic the amplitude structure of
``charming'' penguin contributions~\cite{Ciuchini:2001gv},
charming penguins need not be penguins at
all~\cite{Brodsky:2001yt}.

\section{Calculations of Light-Cone Wavefunctions}

Is there any hope of computing light-front wavefunctions from
first principles?  The solution of the light-front Hamiltonian equation
$ H^{QCD}_{LC}
\ket{\Psi} = M^2 \ket{\Psi}$ is an eigenvalue problem which in principle
determines the masses squared of the entire bound and continuum spectrum
of QCD.  If one introduces periodic or anti-periodic boundary conditions,
the eigenvalue problem is reduced to the diagonalization of a
discrete Hermitian matrix representation of $H^{QCD}_{LC}.$ The
light-front momenta satisfy
$x^+ = {2
\pi
\over L} n_i$ and
$P^+ = {2\pi \over L} K$, where $\sum_i n_i = K.$ The number of quanta in
the contributing Fock states is restricted by the choice of harmonic
resolution.  A cutoff on the invariant mass of the Fock states truncates
the size of the matrix representation in the transverse momenta.
This is
the essence of the DLCQ method \cite{Pauli:1985ps}, which has now become
a standard tool for solving both the spectrum and light-front
wavefunctions
of one-space one-time theories -- virtually any
$1+1$ quantum field theory, including ``reduced QCD"
(which has both quark and
gluonic degrees of freedom) can be completely solved
using
DLCQ \cite{Dalley:1993yy,Antonuccio:1995fs}.
The method yields not only the bound-state and continuum
spectrum, but also the light-front wavefunction
for each eigensolution \cite{Antonuccio:1996hv,Antonuccio:1996rb}.

In the case of theories in 3+1 dimensions, Hiller, McCartor, and I\
 \cite{Brodsky:1998hs,Brodsky:1999xj} have recently shown that the use of
covariant Pauli-Villars regularization with DLCQ allows one to obtain the
spectrum and light-front wavefunctions of simplified theories, such as
(3+1) Yukawa theory.  Dalley \etal\ have shown how one can use DLCQ
in one space-one time, with a transverse lattice to solve mesonic
and gluonic states in $ 3+1$ QCD \cite{Dalley:2000ii}.  The spectrum
obtained for gluonium states
is in remarkable agreement with lattice gauge theory results, but with a
huge reduction of numerical effort.  Hiller and I \cite{Hiller:1999cv}
have shown how one can use DLCQ to compute the electron magnetic moment
in QED without resort to perturbation theory.

One can also formulate DLCQ so
that supersymmetry is exactly preserved in the discrete approximation,
thus combining the power of DLCQ with the beauty of
supersymmetry \cite{Antonuccio:1999ia,Lunin:1999ib,Haney:2000tk}.  The
``SDLCQ" method has been applied to several interesting supersymmetric
theories, to the analysis of zero modes, vacuum degeneracy, massless
states, mass gaps, and theories in higher dimensions, and even tests of
the Maldacena conjecture \cite{Antonuccio:1999ia}.
Broken supersymmetry is
interesting in DLCQ, since it may serve as a method for regulating
non-Abelian theories \cite{Brodsky:1999xj}.

There are also many possibilities for obtaining approximate solutions of
light-front wavefunctions in QCD.  QCD sum rules, lattice gauge theory
moments, and QCD inspired models
such as the bag model, chiral theories, provide important constraints.
Guides to the exact behavior of LC wavefunctions in
QCD can also be obtained from analytic or DLCQ solutions to toy models
such as ``reduced"
$QCD(1+1).$ The light-front and many-body
Schr\"odinger theory formalisms must match In the nonrelativistic limit.

It would be interesting to see if light-front wavefunctions can
incorporate chiral constraints such as soliton (Skyrmion) behavior for
baryons and other consequences of the chiral limit in the soft momentum
regime.  Solvable theories such as $QCD(1+1)$ are also useful for
understanding such phenomena.  It has been shown that the anomaly
contribution for the $\pi^0\to \gamma \gamma$ decay amplitude is
satisfied by the light-front Fock formalism in the limit where the mass
of the pion is light compared to its size \cite{Lepage:1982gd}.

One can also compute the distribution amplitude from the gauge invariant
Bethe-Salpeter wavefunction at equal light-cone time.  This also allows
contact with both QCD sum rules
and lattice
gauge theory; for example, moments of the pion distribution amplitudes
have been computed in lattice
gauge theory \cite{Martinelli:1987si,Daniel:1991ah,DelDebbio:2000mq}.

Dalley \cite{Dalley:2000dh} has recently calculated the pion
distribution amplitude from QCD using a combination of the
discretized DLCQ method for the $x^-$ and $x^+$ light-cone
coordinates with the transverse lattice method
\cite{Bardeen:1976tm,Burkardt:1996gp} in the transverse
directions.  A finite lattice spacing $a$ can be used by choosing
the parameters of the effective theory in a region of
renormalization group stability to respect the required gauge,
Poincar\'e, chiral, and continuum symmetries.  The overall
normalization gives $f_{\pi} = 101$ MeV compared with the
experimental value of $93$ MeV.  The resulting DLCQ/transverse
lattice pion wavefunction with the best fit to the diffractive
di-jet data after corrections for hadronization and experimental
acceptance \cite{Ashery:1999nq}.  The predicted form of
$\phi_\pi(x,Q)$ is somewhat broader than but not inconsistent with the
asymptotic form favored by the measured normalization of $Q^2 F_{\gamma
\pi^0}(Q^2)$ and the pion wavefunction inferred from diffractive di-jet
production.  However, there are
experimental uncertainties from hadronization and theoretical
errors introduced from finite DLCQ resolution, using a nearly
massless pion, ambiguities in setting the factorization scale
$Q^2$, as well as errors in the evolution of the distribution
amplitude from 1 to $10~{\rm GeV}^2$.

Instanton models also
predict a pion distribution amplitude close to the asymptotic form
\cite{Petrov:1999kg}.  In contrast,  recent lattice results from
Del Debbio {\em et al.} \cite{DelDebbio:2000mq} predict a much
narrower shape for the pion distribution amplitude than the
distribution predicted by the transverse lattice.  A new result for
the proton distribution amplitude treating nucleons as chiral
solitons has recently been derived by Diakonov and Petrov
\cite{Diakonov:2000pa}.  Dyson-Schwinger models \cite{Hecht:2000xa}
of hadronic Bethe-Salpeter wavefunctions can also be used to
predict light-cone wavefunctions and hadron distribution
amplitudes by integrating over the relative $k^-$ momentum.  There
is also the possibility of deriving Bethe-Salpeter wavefunctions
within light-cone gauge quantized QCD \cite{Srivastava:2000gi} in
order to properly match to the light-cone gauge Fock state
decomposition.

Clearly much more experimental input on hadron wavefunctions is needed,
particularly from measurements of two-photon exclusive reactions into
meson and baryon pairs at the high luminosity
$B$ factories.  For example, the ratio ${{d\sigma \over dt
}(\gamma
\gamma \to \pi^0
\pi^0)
/ {d\sigma \over dt}(\gamma \gamma \to \pi^+ \pi^-)}$
is particularly sensitive to the shape of pion distribution amplitude.
Baryon pair production in two-photon reactions at threshold may reveal
physics associated with the soliton structure of baryons in
QCD \cite{Sommermann:1992yh}.  In addition, fixed target experiments can
provide much more information on fundamental QCD processes such as deeply
virtual Compton scattering and large angle Compton scattering.

There has been notable progress in computing
light-front wavefunctions directly from the QCD light-front Hamiltonian,
using DLCQ and transverse lattice methods.  Even without full
non-perturbative solutions of QCD, one can envision a
program to construct the light-front wavefunctions using measured moments
constraints from QCD sum rules, lattice gauge theory, and data from
hard exclusive and
inclusive processes.  One can also be guided by theoretical constraints from
perturbation theory which dictate the asymptotic form of the
wavefunctions at large invariant mass,
$x \to 1$, and high
$k_\perp$.  One can also use ladder relations which connect Fock states of
different particle number; perturbatively-motivated numerator spin
structures; conformal symmetry, guidance from toy models
such as ``reduced"
$QCD(1+1)$; and the correspondence to Abelian theory
for
$N_C\to 0$, as well as many-body
Schr\"odinger theory in the nonrelativistic domain.

\section{Calculating and Modelling Light-Cone Wavefunctions}

The discretized light-cone quantization method~\cite{Pauli:1985pv}
is a powerful technique for finding the non-perturbative solutions
of quantum field theories.  The basic method is to diagonalize the
light-cone Hamiltonian in a light-cone Fock basis defined using
periodic boundary conditions in $x^-$ and $x_\perp$.  The method
preserves the frame-independence of the front form.  The DLCQ
method is now used extensively to solve one-space and one-time
theories, including supersymmetric theories.  New applications of
DLCQ to supersymmetric quantum field theories and specific tests
of the Maldacena conjecture have recently been given by Pinsky and
Trittman.  There has been progress in systematically developing
the computation and renormalization methods needed to make DLCQ
viable for QCD in physical spacetime.  For example, John Hiller,
Gary McCartor and I \cite{Brodsky:2001ja} have shown how DLCQ can
be used to solve 3+1 theories despite the large numbers of degrees
of freedom needed to enumerate the Fock basis.  A key feature of
our work, is the introduction of Pauli Villars fields in order to
regulate the UV divergences and perform renormalization while
preserving the frame-independence of the theory.  A review of DLCQ
and its applications is given in Ref. \cite{Brodsky:1998de}.  There
has also been important progress using the transverse lattice,
essentially a combination of DLCQ in 1+1 dimensions together with
a lattice in the transverse dimensions.

Even without explicit solutions, many features of the light-cone
wavefunctions follow from general arguments.  Light-cone
wavefunctions satisfy the equation of motion:
\begin{equation}
H^{QCD}_{LC} \ket{\Psi} = (H^{0}_{LC} + V_{LC} )\ket{\Psi} = M^2
\ket{\Psi}\ ,
\end{equation}
 which has the Heisenberg matrix form
in Fock space:
\begin{equation}
M^2 - \sum_{i=1}^n{m_{\perp i}^2\over x_i} \psi_n =
\sum_{n'}\int \VEV{n|V|n'} \psi_{n'}
\end{equation}
where the convolution and sum is understood over the Fock number,
transverse momenta, plus momenta and helicity of the intermediate
states.  Here $m^2_\perp = m^2 + k^2_\perp.$ Thus, in general,
every light-cone Fock wavefunction has the form:
\begin{equation}
\psi_n={\Gamma_n\over M^2-\sum_{i=1}^n{m_{\perp i}^2\over x_i}}
\end{equation}
 where $\Gamma_n = \sum_{n'}\int V_{n {n'}} \psi_n$.  The main
dynamical dependence of a light-cone wavefunction away from the
extrema is controlled by its light-cone energy denominator.  The
maximum of the wavefunction occurs when the invariant mass of the
partons is minimal; \ie, when all particles have equal rapidity
and are all at rest in the rest frame.  In fact, Dae Sung Hwang
and I \cite{BH2} have noted that one can rewrite the wavefunction
in the form:
\begin{equation} \psi_n= {\Gamma_n\over M^2
[\sum_{i=1}^n {(x_i-{\hat x}_i)^2\over x_i} + \delta^2]}
\end{equation}
where $x_i = {\hat x}_i\equiv{m_{\perp
i}/ \sum_{i=1}^n m_{\perp i}}$ is the condition for minimal
rapidity differences of the constituents.  The key parameter is $
M^2-\sum_{i=1}^n{m_{\perp i}^2/ {\hat x}_i}\equiv -M^2\delta^2.$
We can also interpret $\delta^2 \simeq 2 \epsilon / M $ where $
\epsilon = \sum_{i=1}^n m_{\perp i}-M $ is the effective binding
energy.  This form shows that the wavefunction is a quadratic form
around its maximum, and that the width of the distribution in
$(x_i - \hat x_i)^2$ (where the wavefunction falls to half of its
maximum) is controlled by $x_i \delta^2$ and the transverse
momenta $k_{\perp_i}$.  Note also that the heaviest particles tend
to have the largest $\hat x_i,$ and thus the largest momentum
fraction of the particles in the Fock state, a feature familiar
from the intrinsic charm model.  For example, the $b$ quark has the
largest momentum fraction at small $k_\perp$ in the $B$ meson's
valence light-cone wavefunction,, but the distribution spreads out
to an asymptotically symmetric distribution around $x_b \sim 1/2$
when $k_\perp \gg m^2_b.$

We can also discern some general properties of the numerator of
the light-cone wavefunctions.  $\Gamma_n(x_i, k_{\perp i},
\lambda_i)$.  The transverse momentum dependence of $\Gamma_n$
guarantees $J_z$ conservation for each Fock state: Each light-cone
Fock wavefunction satisfies conservation of the $z$ projection of
angular momentum: $ J^z = \sum^n_{i=1} S^z_i + \sum^{n-1}_{j=1}
l^z_j$. The sum over $s^z_i$ represents the contribution of the
intrinsic spins of the $n$ Fock state constituents.  The sum over
orbital angular momenta $l^z_j = -{\mathrm i}
(k^1_j\frac{\partial}{\partial k^2_j}
-k^2_j\frac{\partial}{\partial k^1_j})$ derives from the $n-1$
relative momenta.  This excludes the contribution to the orbital
angular momentum due to the motion of the center of mass, which is
not an intrinsic property of the hadron \cite{Brodsky:2001ii}. For
example, one of the three light-cone Fock wavefunctions of a $J_z
= +1/2$ lepton in QED perturbation theory is $
\psi^{\uparrow}_{+\frac{1}{2}\, +1} (x,{\vec
k}_{\perp})=-{\sqrt{2}} \frac{(-k^1+{\mathrm i} k^2)}{x(1-x)}\,
\varphi \ ,$ where $ \varphi=\varphi (x,{\vec k}_{\perp})=\frac{
e/\sqrt{1-x}}{M^2-({\vec k}_{\perp}^2+m^2)/x-({\vec
k}_{\perp}^2+\lambda^2)/(1-x)}\ . $ The orbital angular momentum
projection in this case is $\ell^z = -1.$ The spin structure
indicated by perturbative theory provides a template for the
numerator structure of the light-cone wavefunctions even for
composite systems.  The structure of the electron's Fock state in
perturbative QED shows that it is natural to have a negative
contribution from relative orbital angular momentum which balances
the $S_z$ of its photon constituents.  We can also expect a
significant orbital contribution to the proton's $J_z$ since
gluons carry roughly half of the proton's momentum, thus providing
insight into the ``spin crisis" in QCD.

The fall-off the light-cone wavefunctions at large $k_\perp$ and
$x \to 1$ is dictated by QCD perturbation theory since the state
is far-off the light-cone energy shell.  This leads to counting
rule behavior for the quark and gluon distributions at $x \to 1$.
Notice that $x\to 1$ corresponds to $k^z \to -\infty$ for any
constituent with nonzero mass or transverse momentum.

The above discussion suggests that an approximate form for the
hadron light-cone wavefunctions might be constructed through
variational principles and by minimizing the expectation value of
$H^{QCD}_{LC}.$

\section{Structure Functions are Not Parton Distributions}

\begin{figure}[htb]
\centering
\includegraphics[width=3in]{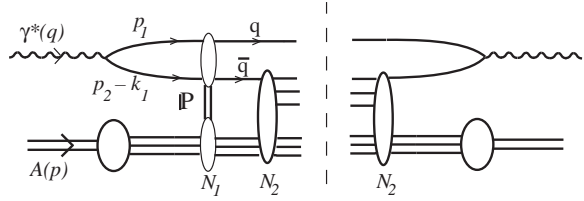}
\caption[*]{Glauber-Gribov shadowing involves interference between
rescattering amplitudes. \label{brodsky2}}
\end{figure}

Ever since the earliest days of the parton model, it has been
assumed that the leading-twist structure functions $F_i(x,Q^2)$
measured in deep inelastic lepton scattering are determined by the
{\it probability} distribution of quarks and gluons as determined
by the light-cone wavefunctions of the target.  For example, the
quark distribution is
\begin{equation}
{ P}_{\qu/N}(x_B,Q^2)= \sum_n \int^{k_{i\perp}^2<Q^2}\left[
\prod_i\, dx_i\, d^2k_{\perp i}\right] |\psi_n(x_i,k_{\perp i})|^2
\sum_{j=q} \delta(x_B-x_j).
\end{equation}
The identification of structure functions with the square of
light-cone wavefunctions is usually made in LC gauge $n\cdot A =
A^+=0$, where the path-ordered exponential in the operator product
for the forward virtual Compton amplitude apparently reduces to
unity.  Thus the deep inelastic lepton scattering cross section
(DIS) appears to be fully determined by the probability
distribution of partons in the target.  However, Paul Hoyer, Nils
Marchal, Stephane Peigne, Francesco Sannino, and I have recently
shown that the leading-twist contribution to DIS is affected by
diffractive rescattering of a quark in the target, a coherent
effect which is not included in the light-cone wavefunctions, even
in light-cone gauge.  The distinction between structure functions
and parton probabilities is already implied by the Glauber-Gribov
picture of nuclear
shadowing~\cite{Gribov:1969jf,Brodsky:1969iz,Brodsky:1990qz,Piller:2000wx}.
In this framework shadowing arises from interference between
complex rescattering amplitudes involving on-shell intermediate
states, as in Fig. \ref{brodsky2}.  In contrast, the wave function
of a stable target is strictly real since it does not have on
energy-shell configurations.  A probabilistic interpretation of
the DIS cross section is thus precluded.

It is well-known that in Feynman and other covariant gauges one
has to evaluate the corrections to the ``handbag" diagram due to
the final state interactions of the struck quark (the line
carrying momentum $p_1$ in Fig. \ref{brodsky2}) with the gauge
field of the target.  In light-cone gauge, this effect also
involves rescattering of a spectator quark, the $p_2$ line in Fig.
2.  The light-cone gauge is singular -- in particular, the gluon
propagator $ d_{LC}^{\mu\nu}(k) =
\frac{i}{k^2+\ieps}\left[-g^{\mu\nu}+\frac{n^\mu k^\nu+ k^\mu
n^\nu}{n\cdot k}\right] \label{lcprop} $ has a pole at $k^+ = 0$
which requires an analytic prescription.  In final-state
scattering involving on-shell intermediate states, the exchanged
momentum $k^+$ is of \order{1/\nu} in the target rest frame, which
enhances the second term in the propagator.  This enhancement
allows rescattering to contribute at leading twist even in LC
gauge.

\begin{figure}[htb]
\includegraphics[width=5in]{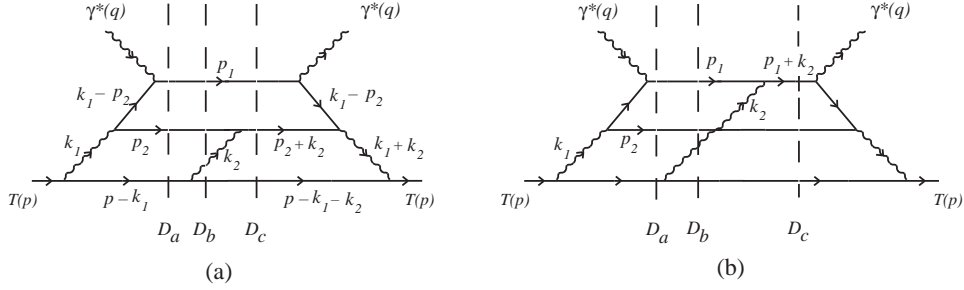}
\caption[*]{Two types of final state interactions.  (a) Scattering
of the antiquark ($p_2$ line), which in the aligned jet kinematics
is part of the target dynamics.  (b) Scattering of the current
quark ($p_1$ line).  For each LC time-ordered diagram, the
potentially on-shell intermediate states -- corresponding to the
zeroes of the denominators $D_a, D_b, D_c$ -- are denoted by
dashed lines. \label{brodsky1}}
\end{figure}

The issues involving final state interactions even occur in the
simple framework of abelian gauge theory with scalar quarks.
Consider a frame with $q^+ < 0$.  We can then distinguish FSI from
ISI using LC time-ordered perturbation
theory~\cite{Lepage:1980fj}.  Figure \ref{brodsky1} illustrates two
LCPTH diagrams which contribute to the forward $\gamma^* T \to
\gamma^* T$ amplitude, where the target $T$ is taken to be a
single quark.  In the aligned jet kinematics the virtual photon
fluctuates into a \qu\qb\ pair with limited transverse momentum,
and the (struck) quark takes nearly all the longitudinal momentum
of the photon.  The initial \qu\ and \qb\ momenta are denoted $p_1$
and $p_2-k_1$, respectively,

The calculation of the rescattering effect of DIS in Feynman and
light-cone gauge through three loops is given in detail in
Ref.~\cite{Brodsky:2001ue}.  The result can be resummed and is
most easily expressed in eikonal form in terms of transverse
distances $r_\perp, R_\perp$ conjugate to $p_{2\perp}, k_\perp$.
The deep inelastic cross section can be expressed as \beq
Q^4\frac{d\sigma}{dQ^2\, dx_B} =
\frac{\alpha}{16\pi^2}\frac{1-y}{y^2} \frac{1}{2M\nu} \int
\frac{dp_2^-}{p_2^-}\,d^2\rvec_\perp\, d^2\Rvec_\perp\,
|\widetilde M|^2 \label{transcross} \eeq where \beq |\widetilde{
M}(p_2^-,\rvec_\perp, \Rvec_\perp)| = \left|\frac{\sin \left[g^2\,
W(\rvec_\perp, \Rvec_\perp)/2\right]}{g^2\, W(\rvec_\perp,
\Rvec_\perp)/2} \widetilde{A}(p_2^-,\rvec_\perp,
\Rvec_\perp)\right| \label{Interference} \eeq is the resummed
result.  The Born amplitude is \beq \widetilde
A(p_2^-,\rvec_\perp, \Rvec_\perp) = 2eg^2 M Q p_2^-\, V(m_\pl
r_\perp) W(\rvec_\perp, \Rvec_\perp) \label{Atildeexpr} \eeq where
$ m_\pl^2 = p_2^-Mx_B + m^2 \label{mplus}$ and \beq V(m\, r_\perp)
\equiv \int \frac{d^2\pvec_\perp}{(2\pi)^2}
\frac{e^{i\rvec_\perp\cdot\pvec_{\perp}}}{p_\perp^2+m^2} =
\frac{1}{2\pi}K_0(m\,r_\perp) \label{Vexpr} \eeq The rescattering
effect of the dipole of the $q \bar q$ is controlled by \beq
W(\rvec_\perp, \Rvec_\perp) \equiv \int
\frac{d^2\kvec_\perp}{(2\pi)^2}
\frac{1-e^{i\rvec_\perp\cdot\kvec_{\perp}}}{k_\perp^2}
e^{i\Rvec_\perp\cdot\kvec_{\perp}} = \frac{1}{2\pi}
\log\left(\frac{|\Rvec_\perp+\rvec_\perp|}{R_\perp} \right).
\label{Wexpr} \eeq The fact that the coefficient of $\widetilde A$
in \eq{Interference} is less than unity for all $\rvec_\perp,
\Rvec_\perp$ shows that the rescattering corrections reduce the
cross section.  It is the analog of nuclear shadowing in our
model.

We have also found the same result for the deep inelastic cross
sections in light-cone gauge.  Three prescriptions for defining
the propagator pole at $k^+ =0$ have been used in the literature:
\beq \label{prescriptions} \frac{1}{k_i^+} \rightarrow
\left[\frac{1}{k_i^+} \right]_{\eta_i} = \left\{
\begin{array}{cc}
k_i^+\left[(k_i^+ -i\eta_i)(k_i^+ +i\eta_i)\right]^{-1} & ({\rm PV}) \\
\left[k_i^+ -i\eta_i\right]^{-1} & ({\rm K}) \\
\left[k_i^+ -i\eta_i \epsilon(k_i^-)\right]^{-1} & ({\rm ML})
\end{array} \right.
\eeq the principal-value, Kovchegov~\cite{Kovchegov:1997pc}, and
Mandelstam-Leibbrandt~\cite{Leibbrandt:1987qv} prescriptions.  The
`sign function' is denoted $\epsilon(x)=\Theta(x)-\Theta(-x)$.
With the PV prescription we have $ I_{\eta} = \int dk_2^+
\left[\frac{1}{k_2^+} \right]_{\eta_2} = 0. $ Since an individual
diagram may contain pole terms $\sim 1/k_i^+$, its value can
depend on the prescription used for light-cone gauge.  However, the
$k_i^+=0$ poles cancel when all diagrams are added;  the net is
thus prescription-independent, and it agrees with the Feynman
gauge result.  It is interesting to note that the diagrams
involving rescattering of the struck quark $p_1$ do not contribute
to the leading-twist structure functions if we use the Kovchegov
prescription to define the light-cone gauge.  In other
prescriptions for light-cone gauge the rescattering of the struck
quark line $p_1$ leads to an infrared divergent phase factor $\exp
i\phi$: \beq \phi = g^2 \, \frac{I_{\eta}-1}{4 \pi} \, K_0(\lambda
R_{\perp}) + {{O}}(g^6) \eeq where $\lambda$ is an infrared
regulator, and $I_{\eta}= 1$ in the $K$ prescription.  The phase is
exactly compensated by an equal and opposite phase from
final-state interactions of line $p_2$.  This irrelevant change of
phase can be understood by the fact that the different
prescriptions are related by a residual gauge transformation
proportional to $\delta(k^+)$ which leaves the light-cone gauge
$A^+ = 0$ condition unaffected.

Diffractive contributions which leave the target intact thus
contribute at leading twist to deep inelastic scattering.  These
contributions do not resolve the quark structure of the target,
and thus they are contributions to structure functions which are
not parton probabilities.  More generally, the rescattering
contributions shadow and modify the observed inelastic
contributions to DIS.

The structure functions measured in deep inelastic
lepton scattering are affected by final-state rescattering, thus
modifying their connection with the light-cone probability
distributions.  In particular, the shadowing of nuclear structure
functions is due to destructive interference effects from
leading-twist diffraction of the virtual photon, physics not
included in the nuclear light-cone wavefunctions.

Our analysis in the $K$ prescription for light-cone gauge
resembles the ``covariant parton model" of Landshoff, Polkinghorne
and Short~\cite{Landshoff:1971ff,Brodsky:1973hm} when interpreted
in the target rest frame.  In this description of small $x$ DIS,
the virtual photon with positive $q^+$ first splits into the pair
$p_1$ and $p_2$.  The aligned quark $p_1$ has no final state
interactions.  However, the antiquark line $p_2$ can interact in
the target with an effective energy $\hat s \propto {k_\perp^2/x}$
while staying close to its mass shell.  Thus at small $x$ and
large $\hat s$, the antiquark $p_2$ line can first multiple
scatter in the target via pomeron and Reggeon exchange, and then
it can finally scatter inelastically or be annihilated.  The DIS
cross section can thus be written as an integral of the
$\sigma_{\bar q p \to X}$ cross section over the $p_2$ virtuality.
In this way, the shadowing of the antiquark in the nucleus
$\sigma_{\bar q A \to X}$ cross section yields the nuclear
shadowing of DIS~\cite{Brodsky:1990qz}.  Our analysis, when
interpreted in frames with $q^+ > 0,$ also supports the color
dipole description of deep inelastic lepton scattering at small
$x$.  Even in the case of the aligned jet configurations, one can
understand DIS as due to the coherent color gauge interactions of
the incoming quark-pair state of the photon interacting first
coherently and finally incoherently in the target.

\newpage

\section{Acknowledgments}
Work supported by the Department of Energy
under contract number DE-AC03-76SF00515.  Much of the new work reported here was
done in collaboration with others, including Susan Gardner, John Hiller,
Dae Sung Hwang, Paul
Hoyer, Nils Marchal, Gary McCartor, Stephane Peigne, Francesco Sannino,
and Prem Srivastava.

\end{document}